\documentclass[aps,prl,twocolumn,notitlepage,superscriptaddress]{revtex4-1}

\usepackage{amsmath}
\usepackage{amsfonts}
\usepackage{amsthm}
\usepackage[utf8]{inputenc}
\usepackage{graphicx}
\usepackage{color}

\usepackage[colorlinks]{hyperref}
\hypersetup{
	bookmarksnumbered,
	pdfstartview={FitH},
	citecolor={blue}, 
	linkcolor={red},
	urlcolor={black},
	pdfpagemode={UseOutlines}
}

\usepackage{upgreek}
\newcommand{\ket}[1]{\left| #1 \right\rangle}
\newcommand{\bra}[1]{\left\langle #1 \right|}

\newcommand{\Ignore}[1]{}

\renewcommand{\eqref}[1]{Eq.~(\ref{#1})}

\newcommand{\mc}[1]{\mathcal{#1}}
\newcommand\abs[1]{\lvert#1\rvert}
\newcommand\Abs[1]{\left|#1\right|}
\newcommand\braXket[4][]{\mathinner{\langle#2\vert#3\vert#4\rangle}_{#1}}

\newcommand{\ox}{\otimes}

\newcommand{\ketbra}[2]{\vert{#1}\rangle\langle{#2}\vert}
 \newcommand{\tr}{\mathrm{Tr}}
 \DeclareMathOperator{\supp}{supp}
\newcommand{\proj}[1]{\ket{#1}\bra{#1}}

\newcommand{\inlineheading}[1]{\textit{{\bfseries#1.---}}}

\newtheorem{Theorem}{Theorem}

\newtheorem{Rem}{Remark}
\newtheorem*{Def}{Definition}

\newtheorem{Prop}{Proposition}
 
\begin{document}

\title{Coherence and quantum correlations measure sensitivity to dephasing channels}

\author{Benjamin Yadin}
\affiliation{$\hbox{Clarendon Laboratory, Department of Physics, University of Oxford, Parks Road, Oxford OX1 3PU, United Kingdom}$}
\author{Pieter Bogaert}
\affiliation{$\hbox{Clarendon Laboratory, Department of Physics, University of Oxford, Parks Road, Oxford OX1 3PU, United Kingdom}$}
\author{Cristian E. Susa}
\affiliation{$\hbox{Departamento de F\'isica y Electr\'onica, Universidad de C\'ordoba, 230002, Monter\'ia, Colombia}$} 
\affiliation{Centre for Bioinformatics and Photonics---CIBioFi, Calle 13 No.~100-00, Edificio 320 No.~1069, Universidad del Valle, 760032 Cali, Colombia}
\author{Davide Girolami}
\email{davegirolami@gmail.com}
\affiliation{$\hbox{Los Alamos National Laboratory, Theoretical Division, Los Alamos 87545, USA}$}

\begin{abstract}
	We introduce measures of quantum coherence as the speed of evolution of a system under decoherence. That is, coherence is the ability to estimate a dephasing channel, quantified by the quantum Fisher information. We extend the analysis to interferometric noise estimation, proving that quantum discord is the minimum sensitivity to local dephasing. A physically motivated set of free operations for discord is proposed. The amount of discord created by strictly incoherent operations is upper-bounded by the initial coherence.
  \end{abstract}

\date{\today}

 
  \maketitle
  
\inlineheading{Introduction}
The peculiar properties of quantum systems can be the source of advantages in quantum information processing tasks \cite{Nielsen2010Quantum}. Quantum resource theories  provide a rigorous means for quantifying quantum properties and understanding their operational utility \cite{Horodecki2013Quantumness,Streltsov2017Colloquium,Chitambar2018Quantum}. Coherence, the superposition of states in a given basis, is arguably the most fundamental quantum trait and has received much recent attention with a resource-theoretic treatment \cite{Streltsov2017Colloquium,Baumgratz2014Quantifying}. A related concept, quantum discord \cite{Henderson2001Classical,Ollivier2001Quantum,Modi2012Classical}, captures the quantumness of correlations between subsystems, and can persist even in the absence of entanglement. Discord does not currently have an associated resource theory, yet it is often quantified in ways similar to coherence. Moreover, a number of formal relations and conversion protocols between the two have been found (see, e.g.\ \cite{Ma2016Converting,Yadin2016Quantum,Adesso2016Measures}). A challenge for the study of both of these resources is to find measures that relate directly to the performance of operational tasks.

Here, we provide an operational interpretation of coherence and discord as the ability of a probe system to encode information about a decoherence process (see Ref.~\cite{epaps} for proofs not in the main text). Namely, both are valuable resources for estimating the strength of a dephasing channel, a metrology primitive which plays a key role in the characterisation of quantum dynamics \cite{est1,est2}, quantum device verification \cite{device}, and tests of fundamental physics, e.g. detection of gravitational effects \cite{gravi}. 
We quantify the utility of a state for this task by the quantum Fisher information (QFI) \cite{Helstrom1969Quantum}. The resulting family of quantities are genuine coherence measures, in particular, being monotonically decreasing under ``strictly incoherent" operations \cite{Winter2016Operational}, which neither create nor use coherence \cite{Yadin2016Quantum}. An interesting consequence of our results is that it is typically impossible to distill a coherent pure state under such operations. We study additional properties of these measures, including an explicit formula for qubits, maximising states, and divergent behaviour. 

Extending to composite systems, we identify discord as the resource for interferometric dephasing estimation. It guarantees the possibility to obtain information about the strength of a dephasing channel on one side of a bipartite system, even when the dephasing basis is unknown. Our proposed measures of discord are derived from the worst-case QFI -- minimised over all local basis choices. They meet a set of consistency criteria, including being valid entanglement measures for pure states, and being monotones under a set of local operations that do not create discord. 
We here define the latter via the physically motivated ``extendibility principle'', advancing towards the sought-after characterisation of the free operations for  discord \cite{Adesso2016Measures}.
It emerges that subtly different resources are at work in interferometry: asymmetry is the resource for unitary perturbations \cite{Bartlett2007Reference,Gour2008Resource,Girolami2013Characterizing,Girolami2014Quantum,Yadin2016General}, while coherence yields sensitivity to non-unitary noise. 
  Finally, we derive an inequality for the conversion of coherence into quantum correlations. This provides an operationally relevant extension of previous results \cite{Streltsov2015Measuring,Ma2016Converting}.

\inlineheading{Measuring coherence}
We consider finite dimensional quantum systems. Given a basis $\{\ket{i}\}_{i=0}^{d-1}$,  quantum systems can exist in states that are not merely probabilistic mixtures of the $\ket{i}$, but coherent superpositions. Naively, the degree of coherence of a state $\rho$  should be related to the size of the off-diagonal elements $\braXket{i}{\rho}{j},\, i \neq j$. Recent studies have formalised this intuition by giving criteria for determining whether a proposed quantity is a genuine measure of coherence \cite{Baumgratz2014Quantifying,Streltsov2017Colloquium}. Valid measures defined so far include distances from $\rho$ to the set of incoherent states, those of the form $\sum_i p_i \proj{i}$, and subtler quantities related to the usefulness of coherent states for phase estimation \cite{Girolami2014Observable} or discrimination \cite{Napoli2016Robustness}.

We observe that coherence also manifests itself as sensitivity to a decoherence process, determining the ability of the system to act as a useful probe of a dephasing channel. A parameter estimation routine consists of three steps \cite{Helstrom1969Quantum}. First, the preparation of a probe system in a certain state. Second, a controlled perturbation imprinting information about the parameter in the probe. Third, a measurement revealing information about the evolved state of the system, which provides an estimate of the parameter  (more complex adaptive strategies are possible) \cite{Helstrom1969Quantum,Giovannetti2011Advances}.  We focus here on the first step, as we are interested in the useful quantum resources contained in the probe state. Let the perturbation inducing decoherence be modelled by a one-parameter, completely-positive trace-preserving (CPTP) map
 \begin{align}
	\Phi^{p}(\rho) & =\rho_p= (1-p)\rho + p \Delta(\rho), \qquad p \in [0,1], \\
	\Delta(\rho) & = \sum_i \ket{i}\bra{i}\rho\ket{i}\bra{i}.  \nonumber
\end{align}
Here, $\Delta$ is the full dephasing channel which removes all off-diagonal elements in the chosen basis $\{\ket{i}\}$.
The parameter $p$ describes the temporal evolution of the dephasing process. For example, in Nuclear Magnetic Resonance (NMR) systems, one has  $p=1-e^{-t/T_2},$ where $t$ is the time parameter and $T_2$ is the transverse relaxation time associated with the qubit \cite{Jones2011Quantum}.

One may wish to experimentally determine the parameter $p$ by observing the time evolution of the system. The accuracy depends on the sensitivity of the state to the dephasing channel $\Phi^p$ -- the faster the evolution, the more precise the estimation.
The instantaneous speed of evolution at time $p$ can be quantified using the quantum Fisher information (QFI)  \footnote{There is a subtlety in the definition of the QFI that becomes relevant at points where the rank of the state changes \cite{Safranek2017Discontinuities}. Our definition coincides with the maximal classical Fisher information obtained from any POVM statistics, thus having a direct connection with performance in metrology. Divergent Fisher information at a parameter value $p$ indicates that $\mathrm{Fid}(\rho_p,\rho_{p+\epsilon})$ has a nonzero first-order term in $\epsilon$, and that the variance of an estimator can be made arbitrarily small in the neighbourhood of $p$.},
\begin{eqnarray}\label{eqn:qfi_defn}
	F(\rho,\Phi^p) :&=& \lim_{\epsilon \to 0} \frac{1- \mathrm{Fid}(\rho_p,\rho_{p+\epsilon})}{8 \epsilon^2} \nonumber\\
	&=&2\sum_{i,j}
	\frac{\left|\bra{\psi^p_i} \partial_p\Phi^{p}(\rho) \ket{\psi^p_j}\right|^2}{\lambda^p_i+\lambda^p_j},
\end{eqnarray}
where $\mathrm{Fid}(\rho,\sigma)$ is the fidelity between two states $\rho,\, \sigma$, and $\Phi^p(\rho) = \sum_i \lambda^p_i \proj{\psi^p_i}$ is the spectral decomposition of the state. A limit to the ability in estimating $p$ is given by the Cram\'er-Rao bound: Suppose that $\mu$ independent copies of $\rho_p$ are used for measurements of an unbiased estimator $O_p$ (such that $\tr O_p \rho_p = p$), then the variance of the estimator is lower bounded by the inverse of the QFI, $\langle (O_p -\langle O_p\rangle)^2\rangle \geq  (\mu F((\rho,\Phi^p))^{-1}$ \cite{Paris2009Quantum}.

We prove that the QFI under dephasing is bona fide measure of coherence, and so denote $C_p(\rho) := F(\rho,\Phi^p)$.
\begin{Theorem}
$C_p(\rho)$ is a valid coherence measure for any $p \in [0,1]$, with respect to a basis $\{\ket{i}\}$, satisfying the criteria \cite{Baumgratz2014Quantifying}:
\begin{enumerate}
	\item[C1)] Faithfulness. Vanishing if and only if the state is incoherent: $C_p(\rho)=0 \Leftrightarrow \rho=\Delta(\rho)$. 
	\item[C2)] Monotonicity under free operations.
	The  set of free operations for coherence is subject of debate \cite{Streltsov2017Colloquium,Marvian2016Quantum,Chitambar2016Critical}. Here we prove monotonicity with respect to the set of strictly incoherent operations (SIOs) \cite{Winter2016Operational,Yadin2016Quantum}.
	They have a physical implementation in interferometric settings \cite{Yadin2016Quantum}. The Kraus operators of an SIO $\mc{E}(\rho) = \sum_k K_k \rho K_k^\dagger$ read $K_k = \sum_i c_{k,i} \ket{f_k(i)}\bra{i}$, where each $f_k$ is a permutation of the set $\{0,1,\dots,d-1\}$.
	
	For any trace-preserving SIO $\mc{E}$ and any state $\rho$, one has $C_p(\mc{E}(\rho)) \leq C_p(\rho)$. When an SIO outputs an ensemble $\sigma_k$ with probabilities $p_k$, one has $\sum_k p_k C_p(\sigma_k) \leq C_p(\rho)$.
	
	\item[C3)] Convexity. For any ensemble of states $\rho_k$ with probabilities $p_k$, one has $C_p(\sum_k q_k \rho_k)\leq \sum_k q_k C_p(\rho_k)$. \\
\end{enumerate}
\end{Theorem}

 As the result holds for any value of $p\in[0,1],$ i.e. each value of $p$ yields a valid  measure, coherence determines the speed of evolution of the system, regardless the values of the  physical parameters under scrutiny, e.g. the relaxation time $T_2$. The coherence of a qubit state with respect to the computational basis $\{\ket{0},\ket{1}\}$ is  measured by the QFI for the phase flip channel $(1-p/2)\rho+p/2(\sigma_z \rho\sigma_z)$  \cite{Collins2013Mixed},
\begin{equation} \label{eqn:coh_qubit}
	C_p(\rho) = \frac{x^2+y^2}{1-(1-p)^2(x^2+y^2)/(1-z^2)},
\end{equation}
in terms of the Bloch representation $\rho = \frac{1}{2}( I + x \sigma_x + y \sigma_y + z \sigma_z)$. Note the special simplifying cases of $p=1$, $C_1(\rho) = R^2 := 4 \abs{\bra{0}\rho \ket{1}}^2$ and of a pure state, $C_p(\proj{\psi}) = R^2 / [p(2-p)]$. Evidently the latter diverges at $p=0$ for any coherent $\ket{\psi}$, signifying the sudden rank change as dephasing is introduced. We generalise these observations to higher dimensions:


\begin{Prop} \label{prop:max_coherence}
For $d$-dimensional systems,	$C_p$ takes the maximal value ${(d-1)/p[d-(d-1)p]}$ on the set of maximally coherent states $\ket{\psi} = \sum_{i=0}^{d-1} \frac{e^{i\theta_i}}{\sqrt{d}} \ket{i}$.
	\begin{proof}
		Due to convexity (C3), 
		the states maximising $C_p$ can be taken to be pure. Next, it is known that a maximally coherent state can be transformed deterministically into any other pure state under SIO \cite{Du2015Conditions,Winter2016Operational}. So the monotonicity condition (C2) shows that no other pure state has a higher value of $C_p$. Considering the state $\ket{\Psi_0}:=\sum_{k=0}^{d-1}\ket{k}/\sqrt{d}$, and the orthonormal set of states $\ket{\Psi_n} := \sum_{k=0}^{d-1} e^{2\pi i n k/d}/\sqrt{d} \ket{k}$, one has $
			\Phi^p(\proj{\Psi_0}) = \left( 1-p+\frac{p}{d} \right) \proj{\Psi_0} + \sum_{n=1}^{d-1} \frac{p}{d} \proj{\Psi_n}.$ This provides the spectral decomposition to be inserted into (\ref{eqn:qfi_defn}).
	\end{proof}
\end{Prop}
As in the qubit case, $C_0$ may diverge:
\begin{Prop}
$C_0(\rho)$ is finite if and only if $\supp \Delta(\rho) = \supp \rho$.
\begin{proof}
	This follows from writing (\ref{eqn:qfi_defn}) at $p=0$ in terms of the spectral decomposition of $\rho = \sum_i \lambda_i \proj{\psi_i},$
$
		C_0(\rho) = \sum_{i,j} \frac{\Abs{\braXket{\psi_i}{[\Delta(\rho)-\rho]}{\psi_j}}^2}{\lambda_i+\lambda_j}.
$
	The quantity is finite if and only if, for all $i,j$ such that $\lambda_i = \lambda_j = 0$, $\braXket{\psi_i}{[\Delta(\rho) - \rho]}{\psi_j} = 0$. This is equivalent to $\braXket{\psi_i}{\Delta(\rho)}{\psi_j}=0$, which says that $\Delta(\rho)$ has a support no larger than that of $\rho$. To obtain the claimed statement, this is combined with the generally true property $\supp \Delta(\rho) \supseteq \supp \rho$.
\end{proof}
\end{Prop}

In the ``typical" case (in the sense of full measure), $C_0$ is finite, while it diverges for any coherent pure state. Combining this with our proof of ensemble monotonicity (C2), we obtain  that it is typically impossible to probabilistically distill a coherent pure state from a mixed state. Indeed, when $\supp \Delta(\rho) = \supp \rho$, there is no {\text SIO} that takes $\rho \to \proj{\psi}$ with nonzero probability, for any coherent $\ket{\psi}$. Note that the result also follows from Theorem 3 in Ref.~\cite{Lami2018Generic}. 
 \begin{figure}[t]
	\includegraphics[scale=1]{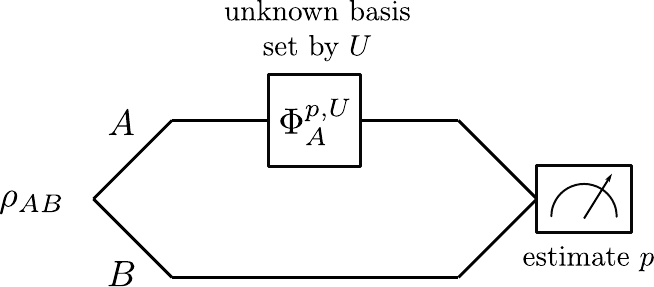}
	\caption{
	 The state $\rho_{AB}$ undergoes a dephasing channel $\Phi^{p,U}_A$ on the side $A$ only, then a joint measurement of both parties is used to estimate the parameter $p$. Even if the dephasing basis (set by the unitary $U$) is unknown, nonzero discord $D_p^A(\rho_{AB})$ guarantees the state being sensitive, conversely to classical correlations.}
	\label{fig:discord_setting}
\end{figure}

\inlineheading{Measuring quantum correlations}
We extend the analysis to show that the sensitivity to any local dephasing  implies the presence of  quantum correlations. A state of a bipartite system $AB$ is called classical-quantum (CQ)   when it takes the form $\rho_{AB} = \sum_i p_i {\proj{\psi_i}}_A \otimes \rho_{B|i}$ for some basis $\{\ket{\psi_i}\}$ on subsystem $A$, and $ p_i\geq 0,\, \sum_i p_i = 1$. By definition, the set of CQ states are those with zero quantum discord  \cite{Ollivier2001Quantum,Henderson2001Classical,Modi2012Classical}. 
We note that $\rho_{AB}$ is CQ if and only if there exists some basis in which dephasing on subsystem $A$ does not perturb $\rho_{AB}$. This motivates defining a measure of discord  as the least sensitivity of a state over all local dephasings,
\begin{align} 
	D_p^A(\rho_{AB}) & := \min_{U \text{ unitary}} \, F(\rho_{AB},\Phi^{p,U}_A), 
\end{align}
where $\Phi^{p,U}(\rho)$ is the dephasing channel with respect to the basis $\{ U \ket{i} \}$, and $\Phi^{p,U}_A := \Phi^{p,U} \otimes I$.

While  a number of criteria identifying valid discord measures have been proposed   \cite{Adesso2016Measures}, we do not have a set of free operations. However,  the full set of local channels (on $A$) which cannot create discord \cite{Hu2012Necessary,Streltsov2011Behavior,Guo2013Necessary}, called commutativity-preserving operations (CPOs), is known. They have the defining property $[\mc{E}(\rho),\mc{E}(\sigma)] = 0$ whenever $[\rho,\sigma]=0$. CPOs are formed by semiclassical, isotropic and unital ($d=2$) channels \cite{Guo2013Necessary} (details in Ref.~\cite{epaps}).
We suggest a principle to constrain this set to a more physical free set: any free operation on  $S$ must admit a dilation in terms of a free operation   on a larger system $SR$. It is often true for resource theories that  whenever $\mc{E}_S$ is free, the trivially extended operation $\mc{E}_S \otimes I_R$ is also free. Applied to CPOs, however, this holds true only for unitary operations. We suggest as a weaker requirement the following:

\begin{Def}[Extendibility principle]
	For any extension $\rho_{SR}$ of a state $\rho_S$ to a larger system, if $\mc{E}_S$ is free then there exists a free operation $\mc{F}_{SR}$ such that $\mc{E}_S(\rho_S) = \tr_R \circ \mc{F}_{SR}(\rho_{SR})$.
\end{Def}

Applied to the local free operations for discord, this results in a set which we name \emph{extendible commutativity-preserving operations} (ECPOs):

\begin{Prop} \label{prop:cond_ECPO}
	A map $\mc{E}$ is an ECPO if and only if it is a semiclassical channel, or an isotropic channel taking the form $\mc{E}(\rho) = t U \rho U^\dagger + (1-t)I/d$ with $t \in [0,1]$.
\end{Prop}

We prove \cite{epaps} that ECPOs consist of either full decoherence in some basis or else a combination of unitary rotations and white noise. Given this result, we obtain:
\begin{Theorem}
	$D_p^A$ is a valid discord measure for any $p \in [0,1]$, satisfying criteria (D1-4):
\begin{enumerate}
	\item[D1)] Faithfulness. Vanishing if and only if the state is CQ: $D_p^A(\rho_{AB}) = 0 \Leftrightarrow \rho $ is CQ.

	\item[D2)] Monotonicity under local operations on the unmeasured subsystem $B$. For any CPTP map $\mc{E}_B$, $D_p^A(\mc{E}_B(\rho_{AB})) \leq D_p^A(\rho_{AB})$.
	
	\item[D3)] Reduction to an entanglement measure for pure states. If $\ket{\psi}$ can be transformed to an ensemble $\{\ket{\phi_k}\}$ with probabilities $q_k$ under local operations and classical communication (LOCC), then the average discord is no larger than the initial discord, $\sum_k q_k D_p^A(\proj{\phi_k}) \leq D_p^A(\proj{\psi})$.
	
	\item[D4)] Monotonicity under local ECPOs on the measured subsystem $A$. For any such $\mc{E}_A$, $D_p^A(\mc{E}_A(\rho_{AB})) \leq D_p^A(\rho_{AB})$.
\end{enumerate}
\end{Theorem}
The measures $D_p^A$ enjoy an operational interpretation, determining the worst-case performance in noise estimation via interferometry (Fig.~\ref{fig:discord_setting}). Suppose that a state $\rho_{AB}$ undergoes a dephasing $\Phi^{p,U}$ on $A$, where both $p$ and $U$ are unknown (i.e. the basis choice is undisclosed), and $p$ is to be estimated.  Then, $D_p^A(\rho_{AB})$ quantifies the worst-case utility of the state for estimating $p$.  In fact, if $\rho_{AB}$ is CQ, the sensitivity to dephasing may be arbitrarily low. \\
We address the question of finding the optimal states for  local dephasing estimation. Firstly, we argue that pure states are optimal. For any $\rho_{AB}$, we find a purification ${\ket{\psi}}_{ABC}$, then it follows from (D2) that $D_p^A(\rho_{AB}) \leq D_p^A({\proj{\psi}}_{ABC})$. Next, due to pure state LOCC monotonicity (D3), this must be maximised by taking ${\ket{\psi}}_{ABC}$ to be maximally entangled. Thus the optimal states are of the form
${\ket{\psi_\mathrm{max}}}_{AB} = \sum_{i=0}^{d-1} \frac{1}{\sqrt{d}} {\ket{i}}_A {\ket{i}}_B, \quad d = \min\{d_A,d_B\}$
for any product basis $\{{\ket{i}}_A {\ket{j}}_B\}$, having compressed $BC$ into a single subsystem $B$. This recovers a result previously obtained in channel estimation via a more convoluted proof \cite{Fujiwara2003Quantum}.
Note that for such states the QFI takes the same value regardless of the chosen basis. This follows from the feature of maximally entangled states that allows a unitary on one side to be transferred onto the other: $U_A \ket{\psi_\mathrm{max}} = U_B^T \ket{\psi_\mathrm{max}}$. By property (D2), such a transformation leaves $D^A_p$ invariant. Thus we can calculate the coherence in the Schmidt basis to obtain the maximal value of ${(d-1)/p[d-(d-1)p]}$, as in Proposition \ref{prop:max_coherence}.

We compare our measures of coherence and discord against related quantities that have appeared and also employ the QFI. Instead of a dephasing channel, one may consider a family of unitary channels $\mc{U}_t(\rho) = e^{-it H} \rho e^{it H}$ for some given Hamiltonian $H$. The QFI with respect to the parameter $t \in \mathbb{R}$ is now a measure of time-translation asymmetry \cite{Bartlett2007Reference} (note that the measure is also independent of $t$). Asymmetry also depends on the existence of coherence in the eigenbasis $\{\ket{i}\}$ of $H$; however, the resulting resource theory has a different structure since it also depends on the eigenvalues of $H$ -- hence, one can treat this asymmetry as a different variety of coherence \cite{Marvian2016Quantum}. The QFI is a monotone under translationally covariant operations, which are a subset of SIOs \cite{Yadin2016General}. The same applies  to 
a more general family of quantities based on monotone metrics \cite{Petz2002Covariance,Zhang2017Detecting,Wigner1963Information,Girolami2014Observable}. Our measures $C_p$ are different in that they are monotones under the greater SIO class of free operations. Similarly, there are measures of discord based on minimising QFI and related quantities with respect to local Hamiltonians, e.g. the \emph{interferometric power} \cite{Girolami2013Characterizing,Girolami2014Quantum}, where the minimisation is over all local  $H_A$ with a fixed spectrum. 
Properties close to (D1-4) have been shown to hold for this measure (although built from a different subset of CPOs) \cite{Bromley2017There}. 
 A recent work \cite{Kim2018Characterizing} introduces a related measure based on unitary QFI, but without a clear operational meaning.
 
\begin{figure}[t]
	\includegraphics[scale=.8]{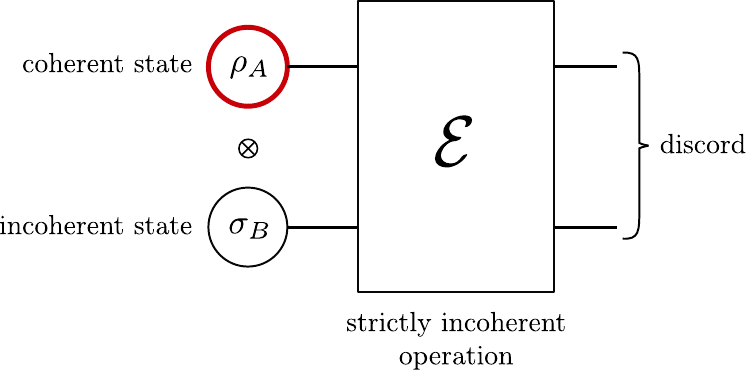}
	\caption{Converting coherence to quantum discord. Two subsystems initially in a product state $\rho_A \ox \sigma_B$, where $\sigma_B$ is incoherent, are coupled by a strictly incoherent channel $\mc{E}$. The discord of the final state is lower-bounded by the initial coherence: $D_p^A(\mc{E}_{AB}[\rho_A \ox \sigma_B]) \leq C_p(\rho_A)$.}
	\label{fig:coherence_to_discord}
\end{figure}
 
We finally discuss the interplay between metrological resources, highlighting how coherence can be traded for quantum correlations. Consider the scenario in Fig.~\ref{fig:coherence_to_discord}.  A probe in an uncorrelated input state $\rho_A \ox \sigma_B$, where $\sigma_B$ is an incoherent state, is subject to  a strictly incoherent operation $\mc{E}_{AB}$ with respect to the product basis $\{{\ket{i}}_A {\ket{j}}_B\}$  (in fact, we only need to assume $[\mc{E}_{AB},\Phi^p_A]=0$). It is interesting to study the properties of the output state   $\mc{E}_{AB}(\rho_A \ox \sigma_B)$ under such a constraint. While a SIO cannot increase coherence, it can convert a basis-dependent resource, the initial coherence, into a basis-independent one, i.e. quantum discord. Property (C2) indeed implies
\begin{align}
	C_p(\rho_A) & = C_p(\rho_A \ox \sigma_B) \nonumber \\
		& \geq C_p(\mc{E}_{AB}(\rho_A \ox \sigma_B)) \nonumber \\
		& \geq D_p^A(\mc{E}_{AB}(\rho_A \ox \sigma_B)).
\end{align}
The result extends to the metrological context previous findings for geometric and entropic measures of coherence and discord
   \cite{Streltsov2015Measuring,Ma2016Converting,Regula2018Converting}.

\inlineheading{Conclusion}
We have showed that quantum coherence can be interpreted as the sensitivity to a decoherence mechanism. While asymmetry determines the speed of evolution of a system under unitary transformations, coherence dictates its dynamics under a dephasing channel.   We have then built a measure of quantum discord as the minimum sensitivity to a local dephasing, showing that local coherence upper bounds the creation of quantum correlations under bipartite strictly incoherent operations.  Dephasing is one of the main sources of noise in quantum information processing, thus being a serious obstacle to develop large scale quantum technologies.   Coherence then yields the usefulness of a probe system for estimating decoherence-related noise, while quantum discord determines the minimal precision in noise estimation via interferometry. 

It would be interesting to  include other quantum resources, e.g.\ entanglement, in such a framework, as well as building metrological measures of genuine multipartite correlations and their complexity \cite{Girolami2017Quantifying}, evaluating a system sensitivity to multilocal dephasings. In fact, asymmetry-related measures such as the interferometric power cannot be straightforwardly constructed in the multipartite case, as quantum correlated states can be unperturbed under unitaries generated by additive many-body Hamiltonians, as these can exhibit degeneracies even if the local terms are non-degenerate. 

\inlineheading{Acknowledgments}
We thank Gerardo Adesso, Fabio Anz\`{a}, Marco Cianciaruso, Bartosz Regula, Luca Rigovacca, Dominik {\v{S}}afr{\'{a}}nek and Tommaso Tufarelli for fruitful discussions. This work was supported by the Los Alamos National Laboratory (project 20180702PRD1), the EPSRC (Doctoral Prize and Grant No. EP/L01405X/1), and Universidad de C\'ordoba (Grant No. CA-097).

\bibliography{cohfish}

  \newpage
  \renewcommand{\bibnumfmt}[1]{[A#1]}

\renewcommand{\citenumfont}[1]{{A#1}}
\setcounter{page}{1}
\setcounter{equation}{0}

\appendix*

\section*{SUPPLEMENTARIES}

\subsection{Proof of Theorem 1} \label{app:coherence}
(C1): $C_p(\rho) = 0$ if and only if $\partial_p \rho_p = 0$, i.e., $\Delta(\rho) - \rho = 0$. \\

(C2): For the sake of clarity, we remind that $C_p(\rho)\equiv F(\rho, \Phi^p)$. The monotonicity of QFI under general quantum channels \cite{Petz2002Covariance} says that
\begin{equation}
	F(\rho, \Phi^p) \geq F(\rho, \mc{E} \circ \Phi^p).
\end{equation}
We use the fact that every SIO $\mc{E}$ is dephasing-covariant \cite{Marvian2016Quantum}, namely $[\mc{E}, \Delta] = 0$. This implies $[\mc{E}, \Phi^p] = 0$, so
\begin{equation} \label{eqn:qfi_monotone}
	F(\rho, \Phi^p) \geq F(\rho, \Phi^p \circ \mc{E}),
\end{equation}
hence $C_p(\rho) \geq C_p (\mc{E}(\rho))$. The ensemble version of monotonicity follows from considering the channel which adds a classical flag to the output, recording a label of the outcome:
\begin{equation}
	\mc{E}(\rho) = \sum_k p_k \proj{k} \otimes \Phi^p(\rho_k).
\end{equation}
The result is found by combining the inequality (\ref{eqn:qfi_monotone}) with the relation $F(\mc{E}(\rho), I \otimes \Phi^p) = \sum_k p_k F(\rho_k, \Phi^p)$. The latter is straightforwardly verified by using the spectral decomposition $\rho_k = \sum_i \lambda_{k,i} \proj{\psi_{k,i}}$ for each $k$. Thus
\begin{widetext}
\begin{align}
	F(\mc{E}(\rho), I \otimes \Phi^p) & = 2 \sum_{i,j,k,l:\, p_k \lambda_{k,i}+p_l\lambda_{l,j} \neq 0} \frac{\abs{\braXket{k \; \psi_{k,i}}{\sum_m p_m \proj{m} \otimes [\Delta(\rho_m) - \rho_m]}{l \; \psi_{l,j}}}^2}{p_k \lambda_{k,i} + p_l \lambda_{l,j}}  \\
		 & = 2 \sum_{i,j,k:\, p_k \lambda_{k,i}+p_k \lambda_{k,j} \neq 0} \frac{p_k^2 \abs{\braXket{\psi_{k,i}}{[\Delta(\rho_k)-\rho_k]}{\psi_{k,i}}}^2}{p_k \lambda_{k,i}+ p_k \lambda_{k,j}} \\
		 & = 2 \sum_k p^2_k \sum_{i,j:\, \lambda_{k,i}+\lambda_{k,j} \neq 0}  \frac{\abs{\braXket{\psi_{k,i}}{[\Delta(\rho_k)-\rho_k]}{\psi_{k,i}}}^2}{p_k \lambda_{k,i}+ p_k \lambda_{k,j}} \\
		 & = \sum_k p_k F(\rho_k,\Phi^p).
\end{align}
\end{widetext}

(C3): This follows directly from convexity of the QFI; alternatively, one can construct the state $\sigma := \sum_k p_k \proj{k} \otimes \rho_k$. As noted above, $F(\sigma,I \otimes \Phi^p) = \sum_k p_k F(\rho_k, \Phi^p)$. Since tracing out a subsystem is an operation which commutes with $\Delta$, it is seen that $F(\sigma, I \otimes \Phi^p) \geq F(\sum_k p_k \rho_k,\Phi^p)$.

\subsection{Proof of Proposition 3} \label{app:free_ops}

The set of CPOs are \cite{Guo2013Necessary}: (i) Semiclassical channels, which always output diagonal states in some basis: $\mc{E}(\rho) = \sum_i p_i(\rho) \proj{i}$. These always destroy discord, i.e., $\mc{E}_A(\rho_{AB})$ is CQ.
(ii) Isotropic channels, of the form $\mc{E}(\rho) = t \Gamma(\rho) + (1-t) I/d$ (in $d$ dimensions), where $\Gamma$ is either unitary or unitarily equivalent to a transpose operation. In the former case, the allowed parameter range is $t \in [\frac{-1}{d^2-1},1]$, in the latter it is $t \in [\frac{-1}{d-1},\frac{1}{d+1}]$.
(iii) In the special case $d=2$, all unital channels: $\mc{E}$ with the property $\mc{E}(I)=I$. 

Now, we prove Proposition \ref{prop:cond_ECPO}.
\begin{proof} 
Firstly, we check that semiclassical channels satisfy the extendibility postulate. Writing $\mc{E}_S(\rho_S) = \sum_i p_i(\rho_S) \proj{i}$, we construct the semiclassical extended channel
\begin{align}
	\mc{F}_{SR}(\rho_{SR}) & := q_{ij}(\rho_{SR}) \proj{ij}, \\
	q_{ij}(\rho_{SR}) & := p_i(\rho_S) \braXket{j}{\rho_R}{j}.
\end{align}
It is easily seen that $\tr_R \circ \mc{F}_{SR}(\rho_{SR}) = \mc{E}_S(\rho_S)$. We must also check that $\mc{F}_{SR}$ is a valid channel. Linearity of $\mc{E}_S$ is equivalent to linearity of the $p_i$, and evidently this implies that $\mc{F}_{SR}$ is linear. Similarly, $\mc{F}_{SR}$ is also trace-preserving. The condition for $\mc{E}_S$ to be completely positive (CP) is the non-negativity of the Choi state \cite[Chapter 8]{Nielsen2010Quantum}:
\begin{align}
	0 & \leq \sum_{j,k} \mc{E}_S(\ket{j}\bra{k}) \otimes \ket{j}\bra{k} \\
		& = \sum_i \proj{i} \otimes \left[ \sum_{j,k} p_i(\ket{j}\bra{k}) \ket{j}\bra{k} \right] \\
	\Leftrightarrow & \sum_{j,k} p_i(\ket{j}\bra{k}) \ket{j}\bra{k} \geq 0 \quad \forall i. \label{eqn:semiclassical_cp}
\end{align}
Similarly, $\mc{F}_{SR}$ is CP when
\begin{align}
	0 & \leq \sum_{k,l,m,n} q_{ij}\big(\ketbra{k}{l} \otimes \ketbra{m}{n}\big) \ketbra{k}{l} \otimes \ketbra{m}{n}\\
		& = \left[ \sum_{k,l} p_i\big(\ket{k}\bra{l} \big) \ket{k}\bra{l} \right] \otimes \proj{j},
\end{align}
which is satisfied thanks to (\ref{eqn:semiclassical_cp}). \\

Next, consider an isotropic channel of the form
\begin{equation}
	\mc{F}_{SR}(\rho_{SR}) = t \rho_{SR} + (1-t) \frac{I_S}{d_S} \ox \frac{I_R}{d_R}, \quad t \in \left[\frac{-1}{d_S^2 d_R^2-1},1 \right].
\end{equation}
Then $\tr_R \circ \mc{F}_{SR}(\rho_{SR}) = t \rho_S + (1-t)I_S / d_S$. By considering arbitrarily large $d_R$, we see that a ``unitary" isotropic channel $\mc{E}(\rho) = t U \rho U^\dagger + (1-t) I/d$ satisfies the extendibility postulate if and only if $t \in [0,1]$. \\

The same argument applied to an ``anti-unitary" isotropic channel $\mc{E}(\rho) = t U \rho^T U^\dagger + (1-t)I/d$ shows that only the trivial case $t=0$ satisfies the extendibility postulate. \\

Finally, it is also clear from above that qubit unital channels are not extendible unless they fall into one of the two classes already allowed.
\end{proof}

\subsection{Proof of Theorem 2} \label{app:discord}
(D1): $D_p^A(\rho_{AB}) = 0$ if and only if there exists a local basis for $A$ in which $\rho_{AB}$ is block-diagonal; this is exactly the condition for $\rho_{AB}$ to be CQ. \\

(D2): Monotonicity under arbitrary operations on $B$ follows immediately from the observation that every operation on $B$ commutes with dephasing on $A$, i.e., $[\mc{E}_B,\Phi^{p,U}_A]=0$, by applying the monotonicity property (C2). \\

(D3): We first note that $D_p^A$ is invariant under local unitaries on $A$, which can be seen from its definition.


For LOCC monotonicity, we use the following fact \cite[Proposition 12.14]{Nielsen2010Quantum}: if $\ket{\psi} \to \{\ket{\phi_k}\}$ with probabilities $q_k$ under LOCC, then there exist unitaries $V_k$ and operators $K_k,\, \sum_k K_k^\dagger K_k = I$ such that
\begin{equation}
	\sqrt{q_k} \ket{\phi_k} = (V_k \ox K_k) \ket{\psi}.
\end{equation}
Then the final average discord is
\begin{align}
	\sum_k q_k D_p^A(\proj{\phi_k}) & = \sum_k q_k D_p^A \left(\frac{1}{q_k} V_k \ox K_k \proj{\psi} V_k^\dagger \ox K_k^\dagger \right) \nonumber \\
		& \hspace{-3em} = \sum_k q_k D_p^A \left( \frac{1}{q_k} I \ox K_k \proj{\psi} I \ox K_k^\dagger \right) \nonumber \\
		& \hspace{-3em}= \sum_k q_k \min_{\{U_k\}} \, F \left(\frac{1}{q_k} I \ox K_k \proj{\psi} I \ox K_k^\dagger, \Phi^{p,U_k}_A \right) \nonumber \\
		& \hspace{-3em} \leq \min_U \sum_k q_k F \left( \frac{1}{q_k} I \ox K_k \proj{\psi} I \ox K_k^\dagger, \Phi^{p,U}_A \right) \nonumber \\
		& \hspace{-3em} \leq \min_U \, F (\proj{\psi}, \Phi^{p,U}_A) \nonumber \\
		& \hspace{-3em} = D_p^A(\proj{\psi}),
\end{align}
where we have used unitary invariance and the property (D2). \\

(D4): We prove this result not just for $D_p^A$ but more generally for a class of discord measures derived from coherence measures:

\begin{Rem}
	Let $C^A(\rho_{AB})$ be a coherence measure which is a monotone under dephasing-covariant operations on $A$, and let
	\begin{equation}
		D^A(\rho_{AB}) := \min_{U\ \textnormal{unitary}} \, C^A(U_A \rho_{AB} U_A^\dagger).
	\end{equation}
	Then $D^A$ is a monotone under local ECPOs on $A$.
	\begin{proof}
		The case of semiclassical channels is trivial, since the output is always CQ. Otherwise we take $\mc{E}(\rho) = t V \rho V^\dagger + (1-t) I/d = V \mc{F}(\rho) V^\dagger$, where $\mc{F} = t \rho + (1-t)I/d$ is a dephasing channel. Now let $U$ be such that $D^A(\rho_{AB}) = C^A(U_A \rho_{AB} U_A^\dagger)$, then
		\begin{align}
			D^A(\mc{E}_A(\rho_{AB})) & = D^A(V_A \mc{F}_A(\rho_{AB}) V_A^\dagger) \nonumber \\
				& \leq C^A( U_A \mc{F}_A(\rho_{AB}) U_A^\dagger) \nonumber \\
				& = C^A(\mc{F}_A(U_A \rho_{AB} U_A^\dagger)) \nonumber \\
				& \leq C^A(U_A \rho_{AB} U_A^\dagger) \nonumber \\
				& = D^A(\rho_{AB}),
		\end{align}
		having used the local unitary invariance of $D^A$, and the fact that $\mc{F}_A$ both commutes with unitaries on $A$ and is an incoherent channel.
	\end{proof}
\end{Rem}

\end{document}